\documentstyle[twoside,fleqn,espcrc2]{article}

\newcommand{\AmS}{{\protect\the\textfont2
  A\kern-.1667em\lower.5ex\hbox{M}\kern-.125emS}}

\def\beq{\begin{equation}}
\def\eeq{\end{equation}}
\def\eq{\end{equation}}

\def\nn{\nonumber}
\def\co#1{{\ifmmode{\cal O}_{#1}\else${\cal O}_{#1}$\fi}}
\def\fr#1.#2.{{#1\over #2}}
\def\FR{\fr}
\def\CR{\cr}
\def\cs#1{{\ifmmode{\cal S}_{#1}\else${\cal S}_{#1}$\fi}}
\def\at{{\ifmmode{\tilde A}\else$\tilde A$\fi}}

\def\EmissT{\not \! \!  E_{T}}

\newcommand{\newc}{\newcommand}

\newc{\eegg}{e^+e^-\gamma\gamma}
\newc{\mmgg}{\mu \mu \gamma\gamma}
\newc{\ttgg}{\tau \tau \gamma\gamma}

\newc{\leplep}{l^+l^-}
\newc{\llgg}{l^+l^- \gamma \gamma}
\newc{\lllgg}{l^+l^-l^{\prime \pm} \gamma \gamma}
\newc{\ljjgg}{l^{\pm}jj \gamma \gamma}
\newc{\lljjgg}{l^+l^-jj \gamma \gamma}

\newc{\eeggE}{ee \gamma \gamma + \EmissT}
\newc{\llggE}{l^+l^- \gamma \gamma + \EmissT}

\newc{\gag}{\gamma\gamma}
\newc{\lgg}{l^{\pm} \gamma \gamma}

\newc{\jjgg}{jj \gamma \gamma}
\newc{\jjjjgg}{4j \gamma \gamma}

\newc{\gsim}{\lower.7ex\hbox{$\;\stackrel{\textstyle>}{\sim}\;$}}
\newc{\lsim}{\lower.7ex\hbox{$\;\stackrel{\textstyle<}{\sim}\;$}}
\newc{\ie}{{\it i.e.}}
\newc{\etal}{{\it et al.}}
\newc{\eg}{{\it e.g.}}
\newc{\kev}{\hbox{\rm\,keV}}

\newc{\tbeta}{\tan\beta}
\newc{\uL}{{\tilde u_L}}
\newc{\uR}{{\tilde u_R}}
\newc{\cL}{{\tilde c_L}}
\newc{\cR}{{\tilde c_R}}
\newc{\tL}{{\tilde t_L}}
\newc{\tR}{{\tilde t_R}}
\newc{\dL}{{\tilde d_L}}
\newc{\dR}{{\tilde d_R}}
\newc{\sL}{{\tilde s_L}}
\newc{\sR}{{\tilde s_R}}
\newc{\bL}{{\tilde b_L}}
\newc{\bR}{{\tilde b_R}}
\newc{\eL}{{\tilde e_L}}
\newc{\eR}{{\tilde e_R}}
\newc{\mhp}{m_{H^\pm}}
\newc{\mhalf}{m_{1/2}}

\newc{\lR}{\tilde{l}_R}
\newc{\lL}{\tilde{l}_L}
\newc{\nL}{\tilde{\nu}_L}
\newc{\na}{\chi^0_1}
\newc{\nb}{\chi^0_2}
\newc{\nc}{\chi^0_3}
\newc{\nd}{\chi^0_4}
\newc{\ca}{\chi^{\pm}_1}
\newc{\cb}{\chi^{\pm}_2}
\newc{\capos}{\chi^{+}_1}
\newc{\caneg}{\chi^{-}_1}

\title{\flushright{\small OHSTPY-HEP-T-96-014} \\ A Global $\chi^2$ Analysis of Electroweak Data (including Fermion Masses and Mixing Angles) in SO(10) SUSY GUTs}

\author{Tom\'{a}\v{s} Bla\v{z}ek,\address{
Department of Physics, The Ohio State University \\ 174 W. 18th Ave., Columbus, OH 43210}
Marcela Carena,\address{Theory Division, CERN\\ Geneva, Switzerland}
Stuart Raby,{\hbox{$^{\rm a}$}}\thanks{Talk presented by S. Raby at the Fourth International Conference on Supersymmetry (SUSY96), College Park, MD, May 29,1996.} and 
Carlos Wagner{\hbox{$^{\rm b}$}}
}

\begin{document}

\begin{abstract}
We present a global $\chi^2$ analysis of electroweak data, including fermion masses and 
mixing angles, in SO(10) SUSY GUTs.  Just as precision electroweak data is used to test
the Standard Model, the well determined Standard Model parameters are the precision
electroweak data for testing theories beyond the Standard Model. In this talk we use the
latest experimentally measured values for these parameters. We study several models
discussed in the literature. One of these models provides an excellent fit to the low energy data
with $\chi^2 \sim 1$ for 3 degrees of freedom.  We present our
predictions for a few selected points in parameter space.  
\end{abstract}
\maketitle

\section{Introduction}

 There are many theories of fermion masses and they are all different.   Some of these
theories assume a grand unified symmetry, while others only require the Standard Model gauge
symmetry.  Some contain low energy supersymmetry, while others do not and some include U(1)
family symmetries to restrict Yukawa matrices; others contain non-abelian family
symmetries. Many theories of fermion masses make predictions only for a restricted 
subset of low energy observables. For example, some test bottom-tau unification, the 
Georgi-Jarlskog relation ${m_s \over m_d} = {1 \over 9}{m_{\mu} \over m_e}$, or the
Harvey-Ramond-Reiss relation  $V_{cb} = \sqrt{{m_c \over m_t}}$.    One thing is certain, all
of these theories {\bf cannot} describe Nature.  Nevertheless, many are consistent with the
low energy data to within an {\em order of magnitude}.  

There are many who believe that, with our present knowledge of field theory, this is the 
best we can do.   We think not.  We are encouraged in our belief by the following guiding
principle:  that the simple observed pattern of fermion masses and mixing angles is not due
to some {\em random} dynamics at an effective cut-off scale $M$ but is instead evidence that a
small set of fermion mass operators are dominant.  This leads to the fundamental hypothesis of
our work --  \begin{enumerate} \item  that a few effective operators at an effective cut-off
scale $M$ (where $M = M_{Planck} \;{\rm or}\; M_{string}$) (or at a GUT scale $M_G$) dominate
the quantitative behavior of fermion masses and mixing angles; and  \item  that a more
fundamental theory, incorporating Planck or stringy dynamics, will generate an effective
field theory below $M$ including these dominant operators.  \end{enumerate}  

With this hypothesis in mind, we are encouraged to find this effective field theory. 
However, in order to make progress, it is clear that {\em order of magnitude} comparisons
with the data are insufficient.   Moreover, since the predictions of any simple theory are
correlated, a {\em significant test} of any theory requires a {\em global fit to all the low
energy data}.  This work makes the first attempt to bring the tests of theories of fermion
masses into conformity with the accuracy of the low energy data.
 
This analysis would not have been feasible ten years ago. The top quark mass and the 
CKM angle $V_{ub}$ had not been measured; while $V_{cb}$ and the weak mixing angle
$sin^2\theta_W$ were measured, but with large error bars.  Today, as a result of both
experimental and theoretical progress, all but one of the 18 parameters of the Standard
Model are known to much better accuracy [the Higgs mass has yet to be measured].  These
parameters are the precision electroweak data for testing theories of fermion masses.

In this talk we present a global $\chi^2$ analysis of precision electroweak data, 
{\em including fermion masses and mixing angles}, within the context of several theories of
fermion masses based on SO(10) SUSY GUTs.  Of course, the analysis we describe can be
applied to any {\em predictive} theory.  

Why look at SO(10) SUSY GUTs?
\begin{itemize}
\item We use SUSY GUTs since they give the simplest explanation for the experimental 
observation that the three gauge couplings appear to meet at a scale of order $10^{16}$
GeV\cite{DRW}.  Moreover, {\em the grand unified symmetry gaurantees that this result is insensitive to radiative corrections}. 

\item We use SO(10) since it provides the simplest explanation for the
observed family structure of the light fermions\cite{SO(10)}.  \end{itemize}

\section{Low energy observables -- experimental values}

Our $\chi^2$ function includes 20 low energy observables.  In addition we have 
incorporated the experimental bounds on sparticle masses into the code as a  penalty in
$\chi^2$; added if one of these bounds is violated.  This gaurantees that we remain in the
experimentally allowed regions of parameter space. Let us now discuss the experimental
observables and their errors.  

We have 
\begin{itemize}

\item 6 parameters associated with the Standard Model gauge and electroweak symmetry 
breaking sectors --- $\alpha_{EM}$, $ \sin^2\theta_W(M_Z)$,  $M_W$, $  M_Z$, $\rho_{new}$, and $ \alpha_s(M_Z)$; 
\item 13 parameters associated with fermion masses and
mixing angles --- $M_t$, $ m_b(m_b)$, $(M_b-M_c)$, $ m_s$, $m_d/m_s$, $Q$, 
$M_{\tau}$, $M_{\mu}$, $M_e$, $V_{us}$, $V_{cb}$, $V_{ub}/V_{cb}$, and $J$; 
and
\item the branching ratio  for $b \rightarrow s \gamma$
\end{itemize}
where $\rho_{new}$ includes the added contribution to the electroweak $\rho$ parameter 
from physics beyond the SM, $Q$ is the Kaplan-Manohar-Leutwyler ellipse
parameter\cite{kml} relating $u, \,d,$ and $s$ quark masses and $J$ is the Jarlskog CP
violating invariant\cite{jarlskog}. 

The experimental values for each observable are given in table 1 with 
the associated experimental or theoretical error $\sigma$.  {\em Note, $\sigma$ is taken to
be either the experimental error or 1/2\%, whichever is larger.}  This is because our
numerical calculations contain theoretical uncertainties from round-off errors alone which
are of order 1/2\%.\footnote{For $\sin^2\theta_W$ we take $\sigma$ to be  1\%.  This is to
account for the additional error resulting from our neglecting SUSY box and vertex
corrections.}  Thus the listed errors for $M_Z,\; M_W,\; \alpha_{EM},\; M_{\tau},\;
M_{\mu},\; M_e$ are dominated by our theoretical uncertainties.   Note, also,  mass
parameters denoted with a capital letter $M$ are defined as pole masses,  while $m_b, \;
m_s,\; m_d,\; m_u$ are defined as the $\overline{MS}$ running  masses at the scale $m_b$ for
the bottom quark or 1 GeV for the three light quarks.

 The Standard Model values for $\sin^2\theta_W(M_Z)$ or $J$ are not in table 
1.  Instead of $\sin^2\theta_W(M_Z)$ we use the Fermi constant $G_{\mu}$,
while instead of $J$ we use the Bag constant $\hat B_K$. 
 
 The Fermi constant receives small corrections from SUSY, while $\sin^2\theta_W(M_Z)$, on 
the other hand, is more sensitive to the SUSY spectrum\cite{sin2thetaw}.  As a consequence, the value of
$\sin^2\theta_W$ varies by a percent or more in different regions of soft SUSY breaking
parameter space.  In our calculations we include only the leading logarithmic corrections
when evaluating $\sin^2\theta_W(M_Z)$ and we neglect the SUSY box and vertex corrections to
$G_{\mu}$.

 Similarly, we test the CP violating parameter $J$ by comparing to the experimental value 
of $\epsilon_K =  (2.26 \pm 0.02)\times10^{-3}$.  The largest theoretical uncertainty
however comes in the value of the QCD bag constant $\hat B_K$.  We thus give the value for 
$\hat B_K$ as obtained from recent lattice calculations\cite{kilcup}.  We then compare this
to a theoretical value of $\hat B_K$ defined as that value needed to agree with $\epsilon_K$
for a given set of fermion masses and mixing angles, assuming only the SM box diagrams.
 
 The experimental value for $\rho_{new}$ is obtained from Langacker's combined fits to the
precision electroweak data, presented at SUSY96\cite{langacker}.
 
For fermion masses and mixing angles we use those combinations of parameters which are 
known to have the least theoretical and/or experimental uncertainties.  For example, while
the bottom and top quark masses are known reasonably well, the charm quark mass is not
known as accurately.  On the other hand, heavy quark effective theory relates the mass
difference $M_b - M_c$ between the bottom and charm quark {\em pole masses} to 5\%
accuracy\cite{hqet}.  We thus use this relation, instead of the charm quark mass itself, to
test the theory.  $M_b$ and $M_c$ are calculated from the $\overline{MS}$ running
masses, $m_b(m_b)$ and $m_c(m_c)$ using two loop QCD threshold corrections.

Similarly, among the three light quarks there is one good relation which severely 
constrains any theory.   This is the Kaplan-Manohar-Leutwyler ellipse given by 
\beq  1 = {1 \over Q^2} {m_s^2 \over m_d^2} + {m_u^2 \over m_d^2}  
\eeq
or 
\beq  Q = {{m_s \over m_d} \over \sqrt{1 - {m_u^2 \over m_d^2}}} \eeq
where  $Q$ is the ellipse parameter.  The experimental value for $Q$ is obtained from a 
weighted average of lattice results and chiral Lagrangian analysis, with
important contributions from the violation of Dashen's theorem\cite{leutwyler}.  Note that
$Q^2$ is free of $O(m_q)$ corrections, but $m_d/m_s$ is not.  Hence $\sigma$ is so much
smaller for $Q^2$ than for $m_d/m_s$.

The remaining parameters are more or less self evident.  We just remark that the central 
value for $V_{cb}$, as well as the error bars, has steadily decreased in the last 5 years,
making it a very significant constraint.  In addition, the value for $V_{ub}/V_{cb}$
changed dramatically in 1992. It changed from approximately $0.15 \pm 0.05$ to its present
value $0.08 \pm 0.02$, where the errors were and continue to be dominated by theoretical
model dependence.  Clearly the systematic uncertainties were large but are now {\em
hopefully} under control.

\section{Low energy observables -- computed values}

In our analysis we consider the minimal supersymmetric standard model defined at a GUT 
scale with, in all cases  (but one), tree level GUT boundary conditions on gauge couplings
and Yukawa matrices.  In this one case, we include an arbitrary parameter, $\epsilon_3$,
which parametrizes the one loop threshold correction to gauge coupling
unification.\footnote{$\epsilon_3$ is calculable in any complete SUSY GUT.  It
is also constrained somewhat by the bounds on the nucleon lifetime\cite{lucas}.}  We also
include 7 soft SUSY breaking parameters ---   an overall scalar mass $m_0$ for squarks and
sleptons, a common gaugino mass $M_{1/2}$, and the parameters $A_0$, $B$ and $\mu$.  In
addition we have allowed for non-universal Higgs masses, $m_{H_u}$ and
$m_{H_d}$.\footnote{Note that if the messenger scale of SUSY breaking is $M_{Planck}$ then
our analysis is not completely self-consistent. In any complete SUSY GUT defined up to an
effective cut-off scale $M > M_G$, the interactions above $M_G$ will renormalize the soft
breaking parameters.  This will, in general, split the degeneracy of squark and slepton
masses at $M_G$ even if they are degenerate at $M$.  On the other hand, bounds on flavor
changing neutral current processes, severely constrain the magnitude of possible splitting.
Thus these corrections must be small. In addition, in theories where SUSY breaking is
mediated by gauge exchanges  with a messenger scale below (but near) $M_G$,  the present
analysis is expected to apply unchanged.  Since in this case squarks and sleptons will be
nearly degenerate at the messenger scale.  The Higgs mases, on the other hand, are probably
dominated by new interactions which also generate a $\mu$ term.  It is thus plausible to
expect the Higgs masses to be split and independent of squark and slepton masses.  The
parameter $A_0$ could also be universal at the messenger scale.}   Thus the number of
arbitrary parameters in the effective GUT includes the 3 gauge parameters, 7 soft SUSY
breaking parameters and $n_y$ Yukawa parameters. The number $n_y$ and the form of the Yukawa
matrices are model dependent. The effective theory between $M_G$ and $M_Z$ is the MSSM. We
use two loop SUSY renormalization group equations [RGE] (in a $\overline{DR}$ renormalization
scheme) for dimensionless parameters and one loop RGE for dimensionful parameters from $M_G$
to $M_Z$.    However we have checked that the corrections to our results obtained by using
two loop RGE for dimensionful parameters are insignificant.

At $M_Z$ we include one loop
corrections to the W and Z masses.  Thus  the W and Z masses are given by the formulae:
\beq  M_W^2 = {1 \over 2} \,g_2^2 \, v^2 +  \delta_W^2  \eeq \beq M_Z^2 = {1 \over 2}
\,({3 \over 5}\,g_1^2 + g_2^2) \, v^2 + \delta_Z^2 \eeq where $\delta_W,\,
\delta_Z$ are the one loop corrections to the pole masses and $g_1,\, g_2$ are
$\overline{DR}$ gauge couplings evaluated at $M_Z$ in the MSSM.  

The Higgs vacuum expectation value $v$ is an implicit function of soft SUSY breaking 
parameters, gauge and Yukawa couplings.   It is determined by self consistently demanding
minimization of the tree level Higgs potential.  The actual value for $v$ is fixed by
minimizing $\chi^2$.

To compare with the $\overline{DR}$ value of $\sin^2\theta_W(M_Z)$ in the MSSM we use the
relation \beq \sin^2\theta_W = {{3 \over 5} \alpha_1 \over {3 \over 5} \alpha_1 + \alpha_2}
\eeq
 evaluated at $M_Z$.
 
 $\alpha_s(M_Z) \equiv \alpha_3(M_Z)$ is evaluated in the $\overline{MS}$ scheme.  
One loop threshold corrections are included in these parameters in order to correctly
account for states with mass greater than $M_Z$.  Finally for $\alpha_{EM}$  we first evaluate $\alpha(M_Z) = \alpha_2(M_Z) \,\sin^2\theta_W(M_Z)$ in the $\overline{MS}$ scheme and then renormalize  $\alpha(M_Z)$ to zero
momentum.  In both cases the $\overline{DR}$ values of $\alpha_i,\; i=1,2,3$ 
are  first changed to $\overline{MS}$ values using the threshold conditions 
 \beq  (1/\alpha_i(M_Z))|_{\overline{MS}} = (1/\alpha_i(M_Z))|_{\overline{DR}} + C_i \eeq  
with $ C_i = {C_2(G_i) \over 12 \pi}$ and $C_2(G_i) = N$ for SU(N) or $= 0$ for U(1).

For fermion masses, we have at tree level
\beq  m_f = \lambda_f \,{ v \over \sqrt{2}} f(\beta) \eeq
where  $\lambda_f$ are the $3\times3$ Yukawa matrices at $M_Z$ and
$f(\beta) = sin\beta (cos\beta)$ for up quarks (down quarks and charged leptons). In addition, at $M_Z$ we
include the leading ( $O(\tan\beta)$) one loop threshold corrections to fermion masses and
mixing angles\cite{bpr}.  

Finally, we compute the theoretical value for $\hat B_K$ needed to fit $\epsilon_K$ and also
the branching ratio for $b \rightarrow s \gamma$\cite{borzumati}.  

 We then form a $\chi^2$ function including the 20 low energy observables ---  6 in the 
gauge sector ($M_W$, $M_Z$, $\alpha_{EM}$, $\sin^2(\theta_W)$, $\alpha_s$, $\rho$), 13 in
the fermion mass sector (9 charged fermion masses and 4 quark mixing angles) and the
branching ratio for $b \rightarrow s \gamma$.\footnote{The $\chi^2$ function also
includes the conditions for extremizing the Higgs potential.}  This $\chi^2$ function is  minimized
self-consistently by iteratively varying the GUT parameters with $m_0, \; M_{1/2}$, and
$\mu$ fixed.  The procedure takes two steps. First the Higgs mass parameters are varied
until the tree level Higgs potential is minimized, for a fixed trial value for $v$ and
$\tan\beta$.  Then $m_{H_u}$ and $m_{H_d}$ along with the rest of the parameters are varied
using the Minuit routine from the CERN library to minimize the $\chi^2$ function.   

In addition to the 20 observables given in table \ref{tab:observables} we have to deal with 
the experimental lower bounds on sparticle masses.   We require that the contribution of the lightest
neutralino (if below the LEP1 threshold) to the invisible Z partial width is bounded by  $\Gamma(Z \rightarrow\; {\rm invisibles}) = (-1.5 \pm 2.7)$ MeV,
as measured at LEP1.  We also
demand that the lightest chargino be heavier than 65 GeV.  Finally we require the
pseudo-scalar Higgs, A, to have mass greater than 30 GeV.\footnote{Note these bounds are on the
tree level running masses.  The state A (along with the other physical Higgs states)  receives
significant corrections to its tree level mass in the regime of large $\tan\beta$ in which we
work.} In order to take these bounds into account (within Minuit) we have added a large
penalty to the $\chi^2$ function whenever one of these bounds is exceeded.

\section{Model 4 of ADHRS}
In this work we have analyzed several models of fermion masses.  We have studied model 4 
of ADHRS\cite{adhrs}.  In this model $n_y = 5$.  The model is defined by the following  4
operators defined in the effective theory at $M_G$.  $A_1,\; A_2$ and $\tilde A$ are adjoint
scalars with vacuum expectation values in the B-L, Y or X (SU(5) invariant) directions of
SO(10), respectively.

\begin{eqnarray}
\co{33} & = & 16_3 \ 10_1 \ 16_3 \\
\co{23} & = & 16_2 \ {A_2\over\at}\  10_1 \ {A_1\over\at}\  16_3 \nn\\
\co{12} & = & 16_1 \ ({\at\over\cs{M}})^3 \ 10_1 \ ({\at\over\cs{M}})^3\ 
16_2 \nn 
\end{eqnarray}

There are six possible choices for the 22 operator; all give the same $0 : 1 : 3$ Clebsch
relation between up quarks, down quarks and charged leptons responsible for the
Georgi-Jarlskog relation\cite{gj}.

The resulting Yukawa matrices at $M_G$ are given by -- 

$$Y_u=\pmatrix{ 0& C & 0 \CR
C & 0 & -\FR1.3. B \CR
0 & -\FR4.3. B & A}$$
$$Y_d=\pmatrix{ 0& -27 C & 0 \CR
-27 C & E e^{i \phi} & \FR1.9. B \CR
0 & -\FR2.9. B & A}$$
$$Y_e=\pmatrix{ 0& -27 C & 0 \CR
-27 C & 3 E e^{i \phi} & B \CR
0 & 2 B & A}$$

The best fits give $\chi^2 = 13$ for 5
degrees of freedom.   Note that the regions with lowest $\chi^2$
contain significant (of order (4 - 6)\%) one loop SUSY threshold corrections to fermion
masses and mixing angles.   Thus these corrections are necessary to improve the agreement of
the model with the data.

 All
low energy observables are fit to within their $2 \sigma$ bounds.  However the fits for 4
observables, ($V_{cb} = 0.0451$, $V_{ub}/V_{cb} = 0.0468$, $\hat B_K = 0.957$ and $Q^{-2} = 0.00172$) lie in the $(1 - 2)
\sigma$ region.  Note that Minuit typically tries to equalize the contribution of all the
observables  to $\chi^2$.  We now argue that these results are true predictions of the
theory.

Consider the first three parameters.  It was shown by Hall and Rasin\cite{hr} that the 
relation \beq {V_{ub} \over V_{cb}} = \sqrt{{\lambda_u \over \lambda_c}} 
\eeq holds for any fermion mass texture in which the 11, 13 and 31 elements of the mass matrices are zero and perturbative diagonalization is permitted.  Note $\lambda_u, \; \lambda_c$ are the up and charm quark Yukawa couplings evaluated at a common renormalization scale.  A typical value for the right-hand side of the equation is  $0.05$ which is too small for the left-hand side by more than 20\%.

  We now show that the fits for $V_{ub}/V_{cb},\; V_{cb}$ and $\hat B_K$ are correlated.
Consider the formula for $\epsilon_K$ given by
\beq
\epsilon_K \approx (V_{cd}\,{V_{ub} \over V_{cb}} \,V_{cb}^2 sin(\xi))\;\hat B_K \times 
({\rm one \; loop \; factors})
\eeq
where the first factor is just the Jarlskog parameter, $J$ defined by the expression -- \begin{eqnarray} J & = &
Im(V_{ud}V_{ub}^*V_{cb}V_{cd}^*) \\
 &= & |V_{cd}| |V_{ub}/V_{cb}| |V_{cb}|^2 \;\sin{\xi} \nn
\label{eq:J}  \end{eqnarray} where $\xi$ is the CP violating phase.
We see that if $V_{ub}/V_{cb}$ is small, then $V_{cb}$ and $\hat B_K$ must be increased to 
compensate.  As a consequence, $V_{cb}$ and $\hat B_K$ are both too large.  Addition of  13
and 31 mass terms can accomodate larger values of $V_{ub}/V_{cb}$ and thus lower values
for $V_{cb}$ and $\hat B_K$.

Now consider the ellipse parameter $Q$.  This parameter is strongly controlled by the 
Georgi-Jarlskog relation\cite{gj}
\beq  {m_s \over m_d} \approx {1 \over 9}{m_{\mu} \over m_e}  \eeq
which is satisfied by model 4.  This is an important zeroth order relation to try to 
satisfy.  However unless there are small calculable corrections to this relation, it leads
to values of $m_s/m_d \sim 25$ and thus values of $Q$ which are too large.  Note, that
introducing 13 and 31 terms in the down quark and charged lepton mass matrices can, in
principle, perturb the zeroth order Georgi-Jarlskog relation.  

Thus the disagreement between model 4 and the data seems significant.  It is unlikely that 
it can be fixed with the inclusion of small threshold corrections to the Yukawa relations
at $M_G$.  Before one adds new operators, however, it is worthwhile to consider the
possibility that perhaps one of the experimental measurements is wrong.  It is interesting
to ask whether the agreement with the data can be significantly improved by removing one
contribution to the $\chi^2$ function; essentially discarding one piece of data\footnote{We
thank M. Barnett for bringing this idea to our attention}.  In order to check this
possibility we have removed several observables from the analysis (one at a time), to see
if this significantly improves the fit.   We find no significant improvements with this procedure.

On the other hand, we have found that we can indeed improve the results by adding one 
operator contributing to the 13 and 31 elements of the Yukawa matrices.  We discuss
this possibility in the next section.  The additional terms are obtained by adding one new
effective mass operator.  Of course there are many possible 13 operators.  In this work we
have not performed a search over all possible 13 operators. Instead we study two 13
operators which are motivated by two complete SO(10) extensions of model 4.

 \section{Model 4(a,b,c) of LR}
 
In this section we analyze two models derived from complete SO(10) SUSY GUTs discussed 
recently by Lucas and S.R.\cite{lucas}.  The models were constructed as simple
extensions of models 4 of ADHRS.  The label (a,b,c) refers to the different possible 22
operators which give identical Clebsch relations for the 22 element of the Yukawa
matrices.  However in the extension to a complete SUSY GUT these different operators lead
to inequivalent theories.  The different theories are defined by the inequivalent U(1)
quantum numbers of the states.  When one demands ``naturalness", i.e.  includes all terms in
the superspace potential consistent with the symmetries of the theory one finds an
additional 13 operator for models 4(a,b,c) given by -- 

\begin{eqnarray} \co{13} = \\ & (a) & 16_1\  ({\at \over
\cs{M}})^3\  10_1\  ({\at A_2\over{\cs{M}}^2})\  16_3 \nn\\ & (b) & 0 \nn\\ & (c) & 16_1
\ ({\at \over \cs{M}})^3 \ 10_1 \ ({A_2\over \cs{M}})\  16_3 \nn \end{eqnarray}

 One of these models (model 4b) is identical to model 4 of ADHRS when states with mass
greater than $M_G$ are integrated out.  The results for this model are thus
identical to those presented in the previous section; hence we will not discuss it further. 
The other two models include one new effective operator at the GUT scale.   The addition of 
this 13 operator introduces two new real parameters in the Yukawa matrices at $M_G$; thus we
have $n_y = 7$.  

Models 4a and 4c differ only by the 13 operator.   For model 4c the resulting Yukawa
matrices are given by --

$$Y_u=\pmatrix{ 0& C & -\FR4.3. D e^{i \delta} \CR
C & 0 & -\FR1.3. B \CR
\FR1.3. D e^{i \delta} & -\FR4.3. B & A}$$
$$Y_d=\pmatrix{ 0& -27 C & \FR2.3. D e^{i \delta} \CR
-27 C & E e^{i \phi} & \FR1.9. B \CR
-9 D e^{i \delta} & -\FR2.9. B & A}$$
$$Y_e=\pmatrix{ 0& -27 C & -54 D e^{i \delta} \CR
-27 C & 3 E e^{i \phi} & B \CR
- D e^{i \delta} & 2 B & A} .$$

We find that model 4a gives a best fit $\chi^2 \sim 4$ for 3 dof, while model 4c gives 
$\chi^2 \leq 1$ for 3 dof.   For model 4c, the preferred regions of soft SUSY
breaking parameter space for $\mu = 80$  GeV correspond to  $M_{1/2} > 220$ GeV with $m_0 >
300$ GeV.  The lower bound on $m_0$ varies slowly with $M_{1/2}$.  In addition as $\mu$ increases the lower bound on $m_0$ increases significantly.  Finally the lower bound on $M_{1/2}$ is determined by LEP limits on
chargino/neutralino masses. 

Note that the one loop SUSY threshold corrections for fermion masses scale as \beq {\mu M_{1/2}
\over m_0^2} \;\;  {\rm or} \;\; {\mu A_0 \over m_0^2} .\eeq  We find that, in this case, the best
fit tends to minimize these corrections, although for $\chi^2 \sim 1$ these are still of
order (4 - 6)\%.   As a result, in the large region with $\chi^2 < 1$ and the SUSY
corrections to fermion masses negligible, the effective number of degrees of freedom is
actually larger than 3, since in this region there are 7 parameters in the Yukawa matrices
determining the 13 low energy observables in the fermion mass sector.

In table 2, we give the computed values for the low 
energy observables in model 4c and the partial contributions to $\chi^2$ for each observable for a particular point in soft SUSY breaking parameter space.    We note that the best fit value for $\alpha_s$ is always on the low side of 0.12.

In table 3, we also include  values for the 
SUSY spectra and the CP violating angles $\alpha,\; \beta$ and $\gamma$, measurable in
neutral B decays.   We note that the values for  $\alpha,\; \beta$ and $\gamma$ do not
significantly change over soft SUSY breaking parameter space.  Hence the predictions for these CP
violating angles are robust; they are not significantly dependent on the values for the soft
SUSY breaking parameters and thus provide a powerful test of the theory.

\section{Conclusions}

In this talk we have presented a global $\chi^2$ analysis of the low energy data (including fermion masses and mixing angles) within the context of a class of predictive SO(10) supersymmetric GUTs.   One of these models in fact provides an excellent fit to the low energy data.   Moveover since our global analysis can be applied to any predictive theory,  it provides a means of comparing the quality of the fits for different models.  By applying such an analysis to all theories of fermion masses, we may hope to find the effective theory of Nature, valid below the Planck (or string) scale.

This research was supported in part by the U.S. Department 
of Energy contract DE-ER-01545-681.  T.B. would like to thank Piotr Chankowski for many useful discussions.

\protect
\begin{table}
\begin{center}
\label{tab:observables}
\begin{tabular}{|c|c|c|}
\hline
Observable & Central value & $\sigma$ \\
 \hline
$M_Z $            &  91.188       & 0.46     \\
$M_W $            &  80.385       & 0.40     \\
$\alpha_{EM}^{-1}$  &  137.04       & 0.69     \\
$G_{\mu} $        &   $0.11664 \times 10^{-4}$  &  $0.0012 \times 10^{-4}$ \\
$\alpha_s(M_Z)$   &  0.118        & 0.006    \\
$\rho_{new}$      & $-0.9 \times 10^{-3}$  &  $2.9 \times 10^{-3} $    \\
\hline
$M_t  $           &  174.46       & 8.1      \\
$m_b(m_b)  $      &    4.26       & 0.11     \\
$M_b - M_c $      &    3.4        & 0.2      \\
$m_s $            &  180          &  50      \\
$m_d/m_s$         &  0.05         &  0.015    \\
$Q^{-2} $         &  0.00203      &  0.00020     \\
$M_{\tau}$        &  1.777        &   0.0089   \\
$M_{\mu} $        & 105.66        &   0.53    \\
$M_e  $           &  0.5110       &   0.0026   \\
$V_{us}$          &  0.2205       &  0.0026     \\
$V_{cb}$          &  0.040        &  0.003     \\
$V_{ub}/V_{cb}$ &  0.08   &   0.02    \\
$\hat B_K $            &  0.8          &    0.1     \\
\hline
$B(b \rightarrow s \gamma)$ &  $0.232\times 10^{-3}$ &  $0.092\times 10^{-3}$    \\
\hline
\end{tabular}
\caption{Experimental observables}
\end{center}
\end{table}

 \protect
\begin{table}
\label{tab:model4cI}
$$\begin{array}{|c|c|c|c|}
\hline
{\rm Observable} &{\rm Computed \;\; value} & {\rm Contribution \;\; to\;}\chi^2 
 \\ \hline
M_Z             &  91.15       & <0.1   \\
M_W             &  80.38       & <0.1   \\
\alpha_{EM}^{-1}  &  137.0       & <0.1 \\
G_{\mu}        &   0.1165 \times 10^{-4}  &  <0.1  \\
\alpha_s(M_Z)   &  0.1130        & 0.69  \\
\rho_{new}      & 0.1806 \times 10^{-3}  &  0.14    \\
\hline
M_t             &  175.0       & <0.1     \\
m_b(m_b)        &    4.292       & <0.1    \\
M_b - M_c       &    3.468      & 0.12  \\
m_s             &  183.7          & <0.1        \\
m_d/m_s &  0.0496         &  <0.1   \\
Q^{-2}          &  0.002046      &  <0.1    \\
M_{\tau}        &  1.776        & <0.1     \\
M_{\mu}         & 105.7       & <0.1     \\
M_e             &  0.5110       &  <0.1    \\
V_{us}          &  0.2205       &  <0.1      \\
V_{cb}          &  0.04074         &  <0.1     \\
V_{ub}/V_{cb} &  0.07395   &   <0.1    \\
\hat B_K             &  0.8109          &   <0.1     \\
\hline
B(b \rightarrow s \gamma) &  0.2444\times 10^{-3} &  < 0.1      \\
\hline
\end{array}$$
\caption{Results for Model 4c for $\mu = 80$
$M_{1/2} = 280$ and $m_0 = 400$.}
\end{table}

\protect
\begin{table}
\label{tab:model4cIpred}
$$\begin{array}{|l|cccc|}
\hline
{\rm Observable} &\multicolumn{4}{|c|}{\rm Predictions} \\
 \hline
\sin 2\alpha           & 0.95       &   &   & \\
\sin 2\beta             & 0.51       & &   &  \\
\sin 2\gamma  &  -0.66     &  &  & \\
\hline
sin^2\theta_W             & 0.2340     &  &   &   \\
\hline
gluino            & 723              &  & &    \\
charginos   &  70      &260 & &   \\
neutralinos      & 52  &  95 & 130 & 260  \\
\hline
up \; squarks             &  470       & 612 & 746 & 769      \\
down \; squarks       &   509       & 550 & 748 & 773     \\
charged\;  sleptons       &    75      & 339 & 422 & 448  \\
sneutrinos             &  329        & & 440 & 441       \\
\hline
\end{array}$$
\caption{More Results for Model 4c. For squark
and slepton masses, the first two columns are for the third family which are significantly split, while the third and fourth columns are mean values for the nearly degenerate states of the second and first families, respectively.}
\end{table}

\end{document}